\documentclass[aps,prb,10pt,twocolumn,superscriptaddress]{revtex4-2}
\usepackage{amsmath}
\usepackage{amssymb}
\usepackage{graphicx}% Include figure files
\usepackage{dcolumn}% Align table columns on decimal point
\usepackage{bm}% bold math
\newcommand{\eis}{Eu$_5$In$_2$Sb$_6$}

\begin{document}

\title{Surface and electronic structure at atomic length scales of the
non-symmorphic antiferromagnet \eis}

\author{M. Victoria Ale Crivillero}
 \affiliation{Max Planck Institute for Chemical Physics of Solids, N\"othnitzer
 Stra\ss{}e 40, 01187 Dresden, Germany}
\author{Sahana R\"o\ss{}ler}
 \affiliation{Max Planck Institute for Chemical Physics of Solids, N\"othnitzer
 Stra\ss{}e 40, 01187 Dresden, Germany}
% \author{\textcolor{blue}{S. Granovsky}}%
%\author{\textcolor{blue}{M. Dörr}}%
% \affiliation{Institut für Festkörperphysik, Technische Universität Dresden,
% D-01062 Dresden, Germany}
 \author{Priscila F. S. Rosa}
\affiliation{Los Alamos National Laboratory, Los Alamos, New Mexico 87545, USA}
\author{J. M\"uller}
 \affiliation{Institute of Physics, Goethe-University Frankfurt, 60438
 Frankfurt(M), Germany}
\author{U. K. R\"o\ss{}ler}
 \affiliation{IFW Dresden, Helmholtzstra\ss{}e 20, 01069 Dresden, Germany}
\author{S. Wirth}
 \email{Steffen.Wirth@cpfs.mpg.de}
 \affiliation{Max Planck Institute for Chemical Physics of Solids, N\"othnitzer
 Stra\ss{}e 40, 01187 Dresden, Germany}
\date{\today}

\begin{abstract}
We performed Scanning Tunneling Microscopy and Spectroscopy (STM/STS)
measurements to investigate the Zintl phase \eis, a non-symmorphic
antiferromagnet. The theoretical prediction of a non-trivial Fermi surface
topology stabilized by the non-symmorphic symmetry motivated our research. On
the cleaved (010) plane, we obtained striped patterns that can be correlated
to the stacking of the [In$_2$Sb$_6$]$^{10-}$ double chains along the
crystallographic $c$ axis. The attempted cleavage along the $a$ axis revealed
a more complex pattern. We combined the STS measurement on non-reconstructed
(010) and (081) surfaces with DFT calculations to further elucidate the
electronic structure of \eis. From our investigations so far, direct
experimental evidence of the predicted topological surface states remains
elusive.
\end{abstract}
\maketitle

\section{Introduction}
\label{sec:intro}
Over the past decade, the search for materials with nontrivial electronic band
topology has attracted considerable attention  \cite{zha19a,tan19}. The
relevance of symmetry considerations has been put in evidence by the discovery
of various novel topological phases. Non-symmorphic space groups, characterized
by the inexistence of an origin that is simultaneously preserved by all
symmetries, offers a suitable route to explore new topological electronic
materials \cite{wan16}. Indeed, there are theoretical predictions of
non-trivial Fermi surface topology stabilized by the non-symmorphic symmetry
\cite{par13}.

\eis\ crystallizes in a non-symmorphic orthorhombic structure (space group 55,
\textit{Pbam}), characterized by infinite [In$_2$Sb$_6$]$^{10-}$ double chains
along the crystallographic $c$ axis \cite{par02,cha15,rad20,ros20}. A schematic
representation of the crystal structure is presented in Fig.\ \ref{FigG.pdf}(a).
The double chains, separated by three non-equivalent Eu$^{2+}$ ion sites, are
constituted by two InSb$_4$ tetrahedra, bridged by Sb$_2$ dumbbell groups,
see Fig.\ \ref{FigG.pdf}(b). The respective lattice parameters are $a$ =
12.510(3)\,\AA, $b$ = 14.584(3)\,\AA, and $c$ = 4.6243(9)\,\AA\ \cite{par02}.
From the chemical point of view, this structure is an example of a ternary
Zintl phase, which is expected to have a precise electron transference and
therefore to show insulating behavior. Interestingly, the isostructural
compound Ba$_5$In$_2$Sb$_6$ (wallpaper group \textit{pgg}) was predicted to
host topological surface states on the (001) surface \cite{wie18}. Very
recent band structure calculations suggested a similarly non-trivial topology
for the (001) surface of the magnetic analog \eis\ \cite{yfxu}, a claim which
calls for experimental verification.

The inclusion of Eu$^{2+}$ ions into the Zintl phase gives rise to magnetism
\cite{ros20,sub16}. The particular arrangement of the 4$f$ localized moments
dictates a complex, presently not fully elucidated, magnetic structure.
At low temperatures, two antiferromagnetic transitions ($T_{\rm N1} \approx
14$~K and $T_{\rm N2} \approx 7$~K) have been reported \cite{ros20}.
%revealed in our magnetization, heat-capacity, and magneto-transport
%measurements, point to such an intricate magnetic structure.
Notably, \eis\ exhibits a negative magnetoresistance (MR) that increases
strongly upon decreasing the temperature below about 15\,$T_{\rm N1}$ and
reaches a colossal MR (CMR) value of $-$99.999\% at 9 T and 15 K, just above
$T_{\rm N1}$ \cite{ros20}. Concomitant with the rapid increase of the MR, an
anomalous Hall effect (AHE) is observed as well as a deviation of the
susceptibility from a Curie-Weiss-type behavior.

The occurrence of CMR in low-carrier density materials containing Eu$^{2+}$
has been linked to the emergence of quasiparticles called magnetic polarons
\cite{kas68,mol01}. Indeed, the existence of local inhomogeneities in the
electronic density of states at the surface of EuB$_6$ has recently been
seen by scanning tunneling microscopy and spectroscopy (STM/STS) and was
interpreted as the localization of charge carriers due to polaron formation
\cite{poh18}. It is, however, important to note that most of these CMR
materials, ranging from EuB$_6$ \cite{sul00} to manganites \cite{sal01},
exhibit ferromagnetic order. In contrast, \eis, together with EuTe
\cite{oli72,sha72} and Eu$_{14}$MnBi$_{11}$ \cite{cha98}, are candidates for
realizing polarons in an antiferromagnet. We note here that magnetic polarons
in antiferromagnets are explored only scarcely, see e.g.\ \cite{mol07} for a
review.

Here, we report on STM/STS measurements in order to obtained local insights
into the electronic structure and surface morphology of the Zintl phase \eis,
a non-symmorphic antiferromagnet. We combine the STS measurements with DFT
calculations to further elucidate the density of states (DOS). This work
presents, to the best of our knowledge, the fist report of the surface
structure of an Eu-based ternary Zintl phase.

\section{Experimental details}
\label{sec:exp}
\begin{figure}[t]
\includegraphics[width=0.45\textwidth]{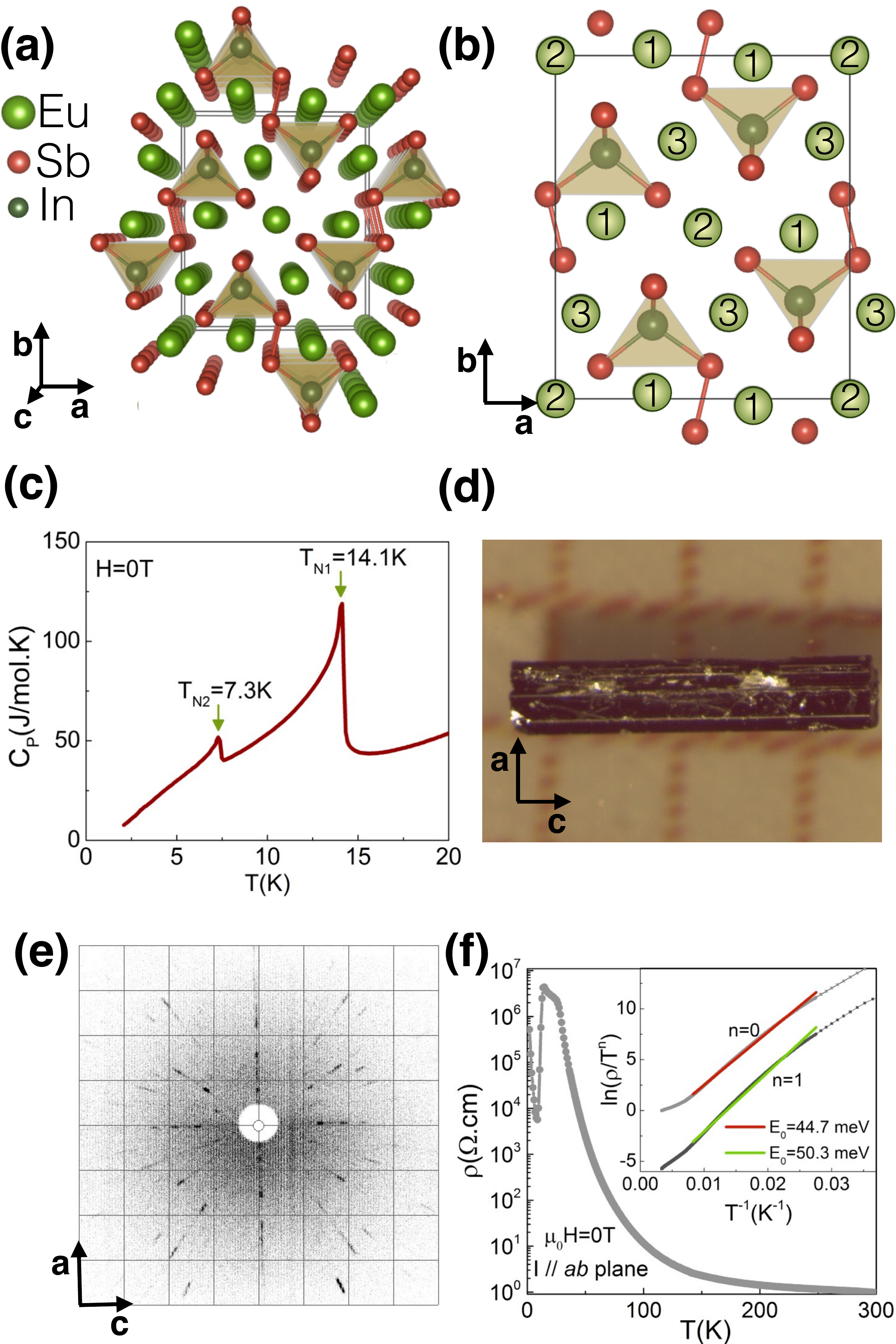}
\caption{Crystal structure of \eis: (a) Perspective projections as viewed
along the [001] axis and (b) cut along the $ab$ plane depicting the three
groups of crystallographically unique Eu atoms, Eu(1), Eu(2), and Eu(3),
labeled respectively as 1, 2, and 3 for simplicity. (c) Specific heat, $C_p$,
as a function of temperature in zero applied magnetic field. (d) Photograph
of an exemplary \eis\ single crystal. (e) X-ray Laue patterns along the
crystallographic $b$ axis. (f) Resistivity $\rho(T)$ as measured in $ab$-plane;
inset: estimates of the activation energy (see text).}
\label{FigG.pdf} \end{figure}

The single crystals investigated in this work were grown using a combined
In-Sb self-flux technique \cite{ros20}. The rod-like shaped crystals have
typical sizes of $0.5$\,mm$\,\times\,0.2$\,mm\,$\times\,3$\,mm, with the
$c$ axis being the long sample axis, see Fig. \ref{FigG.pdf}(d). The
crystallographic structure was verified by X-ray diffraction using a MWL120
real-time back-reflection Laue camera system. Specific heat measurements were
performed using a commercial calorimeter that utilizes a quasi-adiabatic
thermal relaxation technique.

STM/STS was conducted in a commercial low temperature STM (base temperature   $\approx$\,4.6\,K) under ultra-high vacuum (UHV) condition, $p\leq3\times
10^{-9}$\,Pa \cite{Omicron}, using electrochemically etched tungsten tips.
For tunneling spectroscopy, a small ac voltage $V_{\rm mod}$ was added
to the bias voltage $V_b$ and a standard lock-in detection technique was
applied. Some of the reported STM data were obtained with a dual-bias mode:
two different bias voltages were applied for the forward and backward scans.
Therefore, the two topographies were obtained quasi-simultaneously under
otherwise effectively identical conditions, and specifically within identical
areas. If not stated otherwise, the presented STM/STS results were acquired at
base temperature. In addition, we also conducted STM/STS measurements at
elevated temperatures, $5.9$\,K $\leq T\leq 17$\,K, using a resistively
heated sample holder. This allowed us to study the temperature evolution of
the local density of states (LDOS) across both $T_{\rm N1}$ and $T_{\rm N2}$
at atomically resolved length scales.

We attempted to cleave a total of 8 single-crystals \textit{in situ} and at
low temperatures ($T\approx20$\,K) along the main macroscopic sample axes.
As inferred from the diffraction results, those directions are expected to
conform with the crystallographic $a$, $b$ and $c$ axes. However, as a result
of the subtle anisotropy in the \textit{ab} plane, in some cases further
analysis will be needed to unambiguously verify the crystallographic in-plane
sample orientation. In Table~\ref{table:1} details of five surfaces are
presented which were studied in depth. On the remaining samples, atomically
\begin{table}[t]
\caption{Details of the \eis\ single-crystal surfaces studied by STM/STS.}
\begin{ruledtabular}
\begin{tabular}{l|ccccc}
Sample & \#1 & \#2 & \#3 & \#4 & \#5 \\ \hline
Attempted plane& $ac$ & $bc$ & $ac$ & $ab$ & $ab$\\
Assigned plane&(081)& ? & $ac$ & $ab$ & $ab$  \\
surface quality & many & / & extended & \multicolumn{2}{c}{atomically}\\
 & defects & & terraces & \multicolumn{2}{c}{rough}\\
\end{tabular} \end{ruledtabular}
\label{table:1}
\end{table}
flat surface areas were not found.

The electronic structure of \eis\ was also examined by calculations using
density functional theory (DFT). The calculations were performed using the
full-potential local orbital (FPLO) approach \cite{koe99} as implemented in
the fplo code \cite{FPLO}. We used the generalized gradient approximation
(GGA) as exchange-correlation functional \cite{per96}. The magnetic 4$f$ configuration of  Eu$^{2+}$ is expected to remain in the $L=0$ and $S=7/2$ ground-state in \eis. Therefore, we applied the open core approach, which
places a correspondingly occupied 4$f$ shell on the Eu site---in this case an
isotropic shell with a net spin of $7/2$. This entirely removes the 4$f$
levels from the valence band structure \cite{bro91,bro91b,ric98}, while
keeping 4$f$-5$d$ exchange interactions on site and, thus, allows for indirect
exchange couplings via the valence electrons between the Eu-spin moments. Such
a fully localized limit for the 4$f$ electronic states on Eu$^{2+}$ enabled us
to impose various magnetic configurations on the complex crystal structure of
\eis\ and to estimate their energies in a simplified fashion. Within this
method, we explored various collinear spin structures in an effort to examine
possible magnetic ground states of \eis. These calculations of different
magnetic configurations were conducted within the scalar relativistic
approach. In addition, band structures were also evaluated by fully
relativistic calculations. Here, the fully relativistic FPLO code includes
spin orbit coupling (SOC) to all orders, being based on solutions of the
4-spinor Kohn-Sham-Dirac equations.

The calculations of the electronic band-structure of \eis\ were based on the
experimental lattice structure as determined in Ref.~\onlinecite{par02}. For
comparison, we also calculated the electronic band-structure for the
non-magnetic analogue Ba$_{5}$In$_{2}$Sb$_{6}$ for which experimental
structure data were taken from Ref.~\onlinecite{cor88}.

\section{Results}
\subsection{Characterization of \eis\ single crystals}
\label{sec:Charac}
Figure \ref{FigG.pdf}(c) shows the temperature dependence of the specific
heat, $C_p$, in zero magnetic field of a representative \eis\ single crystal.
The overall behavior and more specifically the two antiferromagnetic
\begin{figure}[t]
\includegraphics[width=0.45\textwidth]{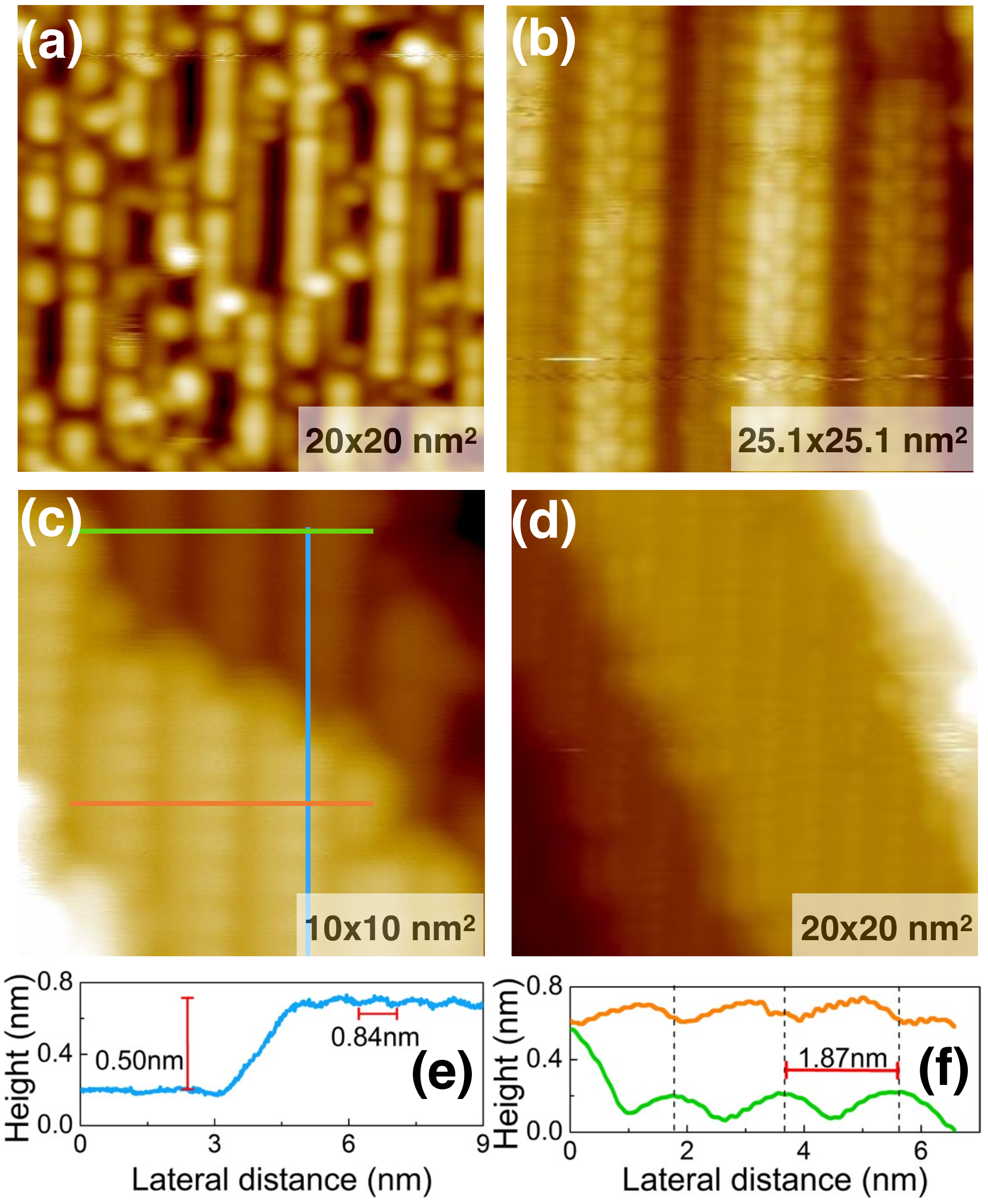}
\caption{Overview of different topographies encountered on \eis\ cleaved
surfaces ($V_b = +2$\,V and current set-point $I_{sp} = 1.4$\,nA). (a), (b)
and (d): Three different areas on sample \#2, $T=4.6$\,K. (c) Topography on
sample \#1 at $T=5.9$\,K and, (e), (f), height profiles along the lines marked
by similar colors in (c).}  \label{FigA.pdf}
\end{figure}
transitions at $T_{\rm N1} = 14.1$\,K and $T_{\rm N1} = 7.3$\,K are in good
agreement with those previously reported \cite{ros20}. The $H$--$T$ phase
diagrams of \eis\ along the three main crystallographic axes, constructed
using a combination of magnetization and heat capacity measurements, will be
reported elsewhere.

Figure \ref{FigG.pdf}(e) presents a representative x-ray Laue pattern along
the crystallographic $b$ axis. For all the single-crystals inspected by x-ray
the main macroscopic sample axes conform with the crystallographic $a$, $b$,
and $c$ axes, with the shortest sample side parallel to the $b$ axis.

\eis\ has been classified as a narrow-gap semiconductor
\begin{figure}[t]
\includegraphics[width=0.45\textwidth]{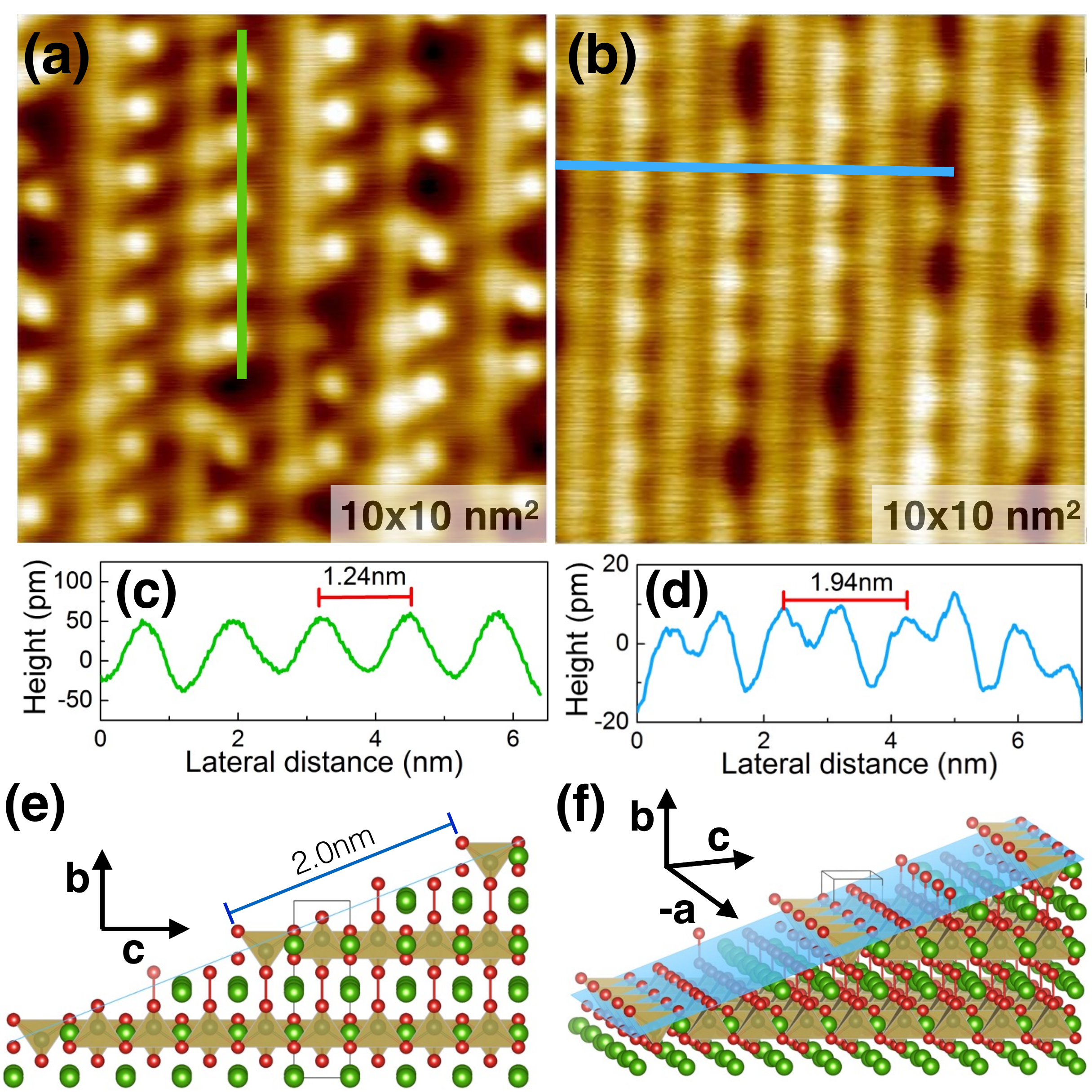}
\caption{STM topography of an atomically flat area of 10$\,\times\,$10\,nm$^2$
on sample \#1, obtained in dual-bias mode at $T = 5.9$\,K and $I_{sp} =
1.4$\,nA. (a) $V_b = +2$\,V; (b) $V_b = -2$\,V. (c) and (d): Height profiles
along the marked lines in (a) and (b), respectively. (e) Schematic
representation of a proposed surface termination cleaved along the plane (081).
Here, the horizontal separation $\Delta x = 2.0$\,nm between the stripes is
marked. (f) Perspective view: the exposed tetrahedra constituting the stripes
along the $a$-axis are separated by 1.25\,nm, in agreement with the measurement
in (c).}  \label{FigE.pdf}
\end{figure}
\cite{ros20,cha15,par02}. This can be inferred from the bulk resistivity
$\rho(T)$, see Fig.\ \ref{FigG.pdf}(f). From fits to $\rho(T) \propto
T^n \exp (-E_a/k_B T)$, activation energies $E_a$ between 45--50 meV were
estimated (for 36\,K $\lesssim T \lesssim$ 120\,K) depending on whether an
Arrhenius ($n = 0$) or small polaron hopping conduction mechanism ($n = 1$)
\cite{emi69} was assumed [inset to Fig.\ \ref{FigG.pdf}(f)]. These values
of $E_a$ are in excellent agreement with those of Ref.\ \onlinecite{ros20}
while much larger values were reported in Ref.\ \onlinecite{cha15}.

\subsection{STM: Topography}
\label{sec:STM}
Interestingly, we found stable tunneling conditions only using relatively
\begin{figure*}
\includegraphics[width=\textwidth]{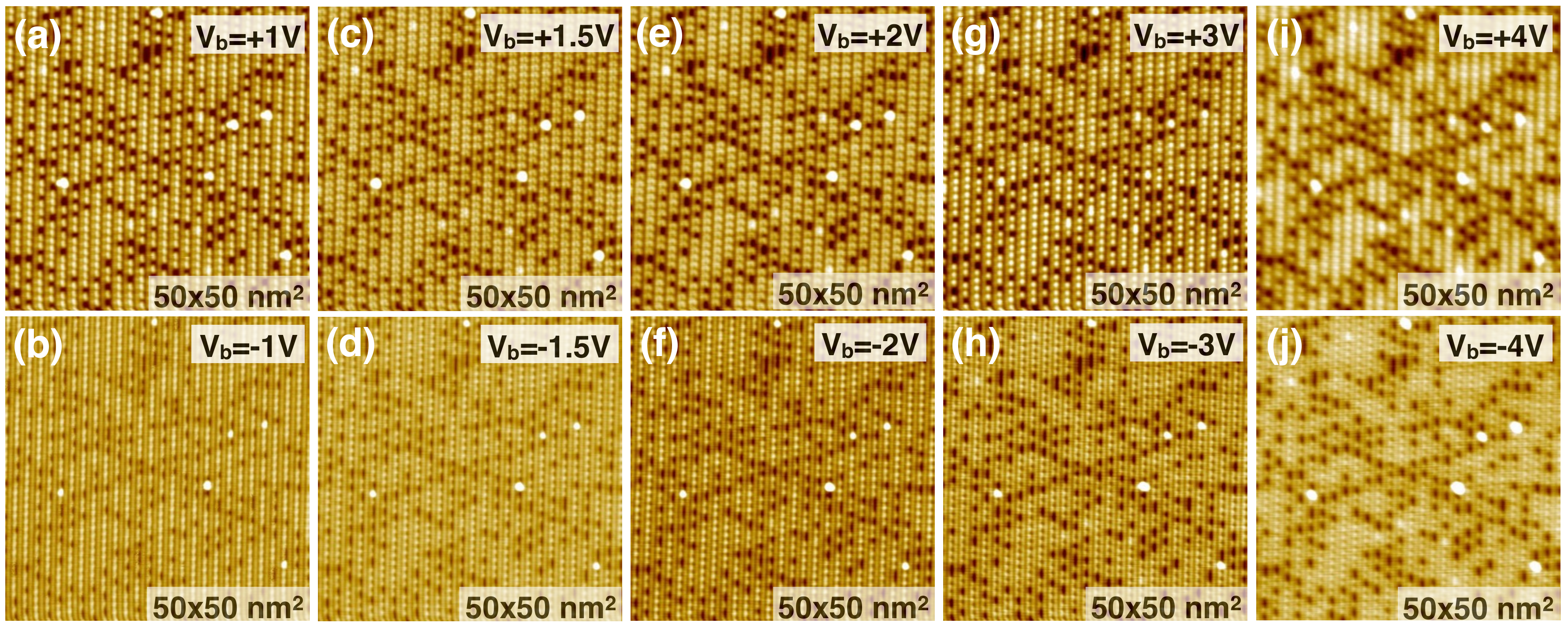}
\caption{STM topographies obtained in dual-bias mode on a relatively big
identical area of $50\times50$\,nm$^2$ on sample \#1 ($I_{sp}=1.4$\,nA,
$T=5.9$\,K). Images obtained within the same run are stacked on top of each
other and with $V_b$ indicated. The color codes of all shown topographies
correspond to a total height scale of approximately 300 pm from black to
white.} \label{FigB.pdf}
\end{figure*}
large bias voltages $V_b$. In Fig.\ \ref{FigA.pdf} an overview of different
atomically flat surfaces encountered on samples \#1 and \#2 is presented
($V_b = +2$\,V). These samples were attempted to be cleaved along the nominal
$a$ axis, i.e. the $bc$ plane is expected to be exposed. In general, this kind
of topography has to be searched for; as mentioned above, on some cleaved
sample surfaces we did not succeed to find atomically flat areas at all. One
principal axis can easily be identified in most cases, the orientation of
which corresponds to the crystallographic $c$ axis. The types of observed topographies ranges from more disordered ones, Fig.\ \ref{FigA.pdf}(a), to
almost defect-free surfaces with step edges, Fig.\ \ref{FigA.pdf}(d). Clearly,
the diversity of observed topographies puts in evidence that \eis\ surfaces
are prone to manifest complex configurations, which reinforces the
relevance of studying the surface properties with a local probe as STM/STS.

The two representative topographies in Figs.\ \ref{FigA.pdf}(c) and (d)
exhibit atomically flat areas separated by sub-nanometer-high step edges.
They were observed on two different samples, \#1 and \#2, over areas of
$10\times10$\,nm$^2$ and $20\times20$\,nm$^2$, respectively. The average
height of the step edges is $0.50$\,nm, as seen from the line scan in Fig.\
\ref{FigA.pdf}(e) taken along the turquoise line in Fig.\ \ref{FigA.pdf}(e).
The distance between corrugations along the $c$-axis in the upper plane is
about $0.84$\,nm. Noticeably, there is a displacement between the corrugations
within the two adjacent planes; compare, for example, the orange and green
line scans in Fig.\ \ref{FigA.pdf}(f). In this case, the distances
between corrugations ($\Delta x \approx 1.87$\,nm, $\Delta y \approx 0.84$\,nm)
and the step height $\Delta z \approx 0.50$\,nm did not match with the
expected values for a non-reconstructed $bc$ plane.

On sample \#1, a different type of atomically flat pattern could be observed,
see Fig.\ \ref{FigE.pdf}. Again, the topographies in Figs.\ \ref{FigE.pdf}(a)
and (b) were obtained in dual-bias mode, i.e. with $V_b = +2$\,V and $-2$\,V,
respectively. In this case, the double-stripe-like corrugations are separated
by $\Delta y \approx 1.94$\,nm, while the distance of corrugations along the
stripe direction is $\Delta x \approx 1.24$\,nm. The closeness of latter
distance to the lattice parameter $a$ suggests that the cleaved surface may
contain the crystallographic $a$ axis. To further identify the surface plane,
we searched in the crystal structure for planes with an orientation close to
$ac$ and which contain features separated by 1.94\,nm. A suitable candidate is
the (081) plane; see the schematic representations in Figs.\ \ref{FigE.pdf}(e)
and (f). In that case, the distance between the exposed tetrahedra along the
$a$-axis is $a$, while the horizontal separation between the so-formed
stripes is 2.00\,nm. The tilting angle between the (081) plane and the $ac$
plane is $22^{\circ}$. Noticeably, some crystals exhibit facets at
$\sim$30$^{\circ}$ from the $a$-axis. Therefore, a plausible scenario is that
this sample was mounted parallel to one of those facets before cleaving
such that the crystal was then cleaved along the (081) plane. Regarding the
topographies Fig.\ \ref{FigE.pdf}(a) and (b), it is worth noting three
observations. First, there is a noticeable difference in contrast for the two
$V_b$-values applied (see also Fig.\ \ref{FigB.pdf}) suggesting a complex
energy and spatial dependence of the types of states dominating the tunnel
current. Second, this type of pattern extends over relatively large areas
without step edges. Third, there is an appreciable number of defects observed.

To gain further insight into the dependence of the contrast in topography on
$V_b$, a series of 5 images of the same area was taken in dual-bias mode
(positive and negative $V_b$) with $|V_b|$ increasing from 1--4 V, see Fig.\
\ref{FigB.pdf}. Note that these images were obtained on the same cleave and
close to the area shown in Figs.\ \ref{FigE.pdf}(a),\,(b). While the contrast
is strongly enhanced for positive $V_b$, there is no contrast reversal or
lateral shift of corrugations observed upon reversing $V_b$, neither for the
regular features nor for the defects. These observations are consistent with
an above-indicated assignment of the regularly arranged protrusions to
\begin{figure}[t]
\includegraphics[width=0.45\textwidth]{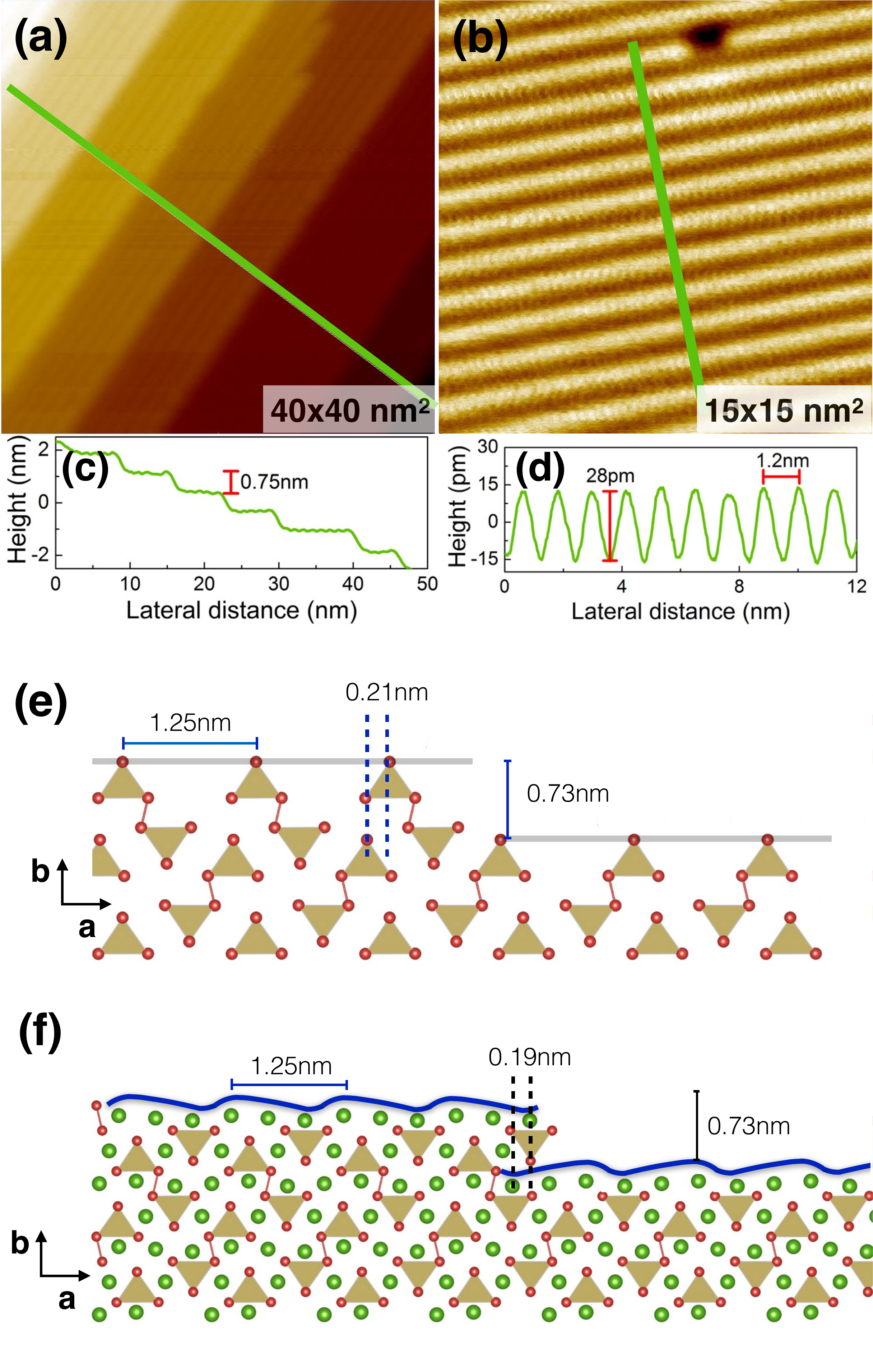}
\caption{Topography of non-reconstructed, atomically flat surfaces
cleaved along the $ac$ plane (sample \#3). (a) and (b): STM topographies
obtained with $V_b = -3$\,V and $I_{sp} = 1.4$\,nA at $T = 7.0$\,K and $T =
5.9$\,K, respectively. (c) and (d): Height profiles along the corresponding
lines marked in (a) and (b). (e) Schematic representation of the proposed Sb-terminated surface cleaved along the $ac$ plane and a corresponding
sub-unit cell terrace. Note that for simplicity only the tetrahedra are shown.
Their double-chain stacking imposes a lateral spacing of the protrusions of
$a=1.25$\,nm along the $a$-axis and a vertical distance to the next
Sb-terminated surface $\frac{1}{2}b = 0.73$\,nm. (f) Alternative Eu-terminated
(010) plane formed by breaking the Sb dumbbell bonds during the cleaving.
The resulting height-modulated surface also exhibits a 1.25\,nm period along
$a$ and 0.73\,nm step edges along $b$.}  \label{FigC.pdf}
\end{figure}
negatively charged Sb in either the InSb$_4$ tetrahedra or the Sb$_2$
dumbbells, cf.\ Figs.\ \ref{FigE.pdf}(e),\,(f). However, to fully interpret
the STM topographies, a comparison with simulated STM images will be required
\cite{mon13,woo12}.

Motivated by the results presented above, the following sample \#3 was
attempted to be cleaved along the $ac$ plane. Here, we were able to locate some
areas where the in-plane arrangement of corrugations and the height of step
edges match the values expected for an $ac$ surface. In Fig.\ \ref{FigC.pdf}
we present a striped pattern with a lateral spacing of $1.2$\,nm, while the
height difference between terraces is $0.75$\,nm $\approx\frac{1}{2}b$.
\begin{figure}
\includegraphics[width=0.45\textwidth]{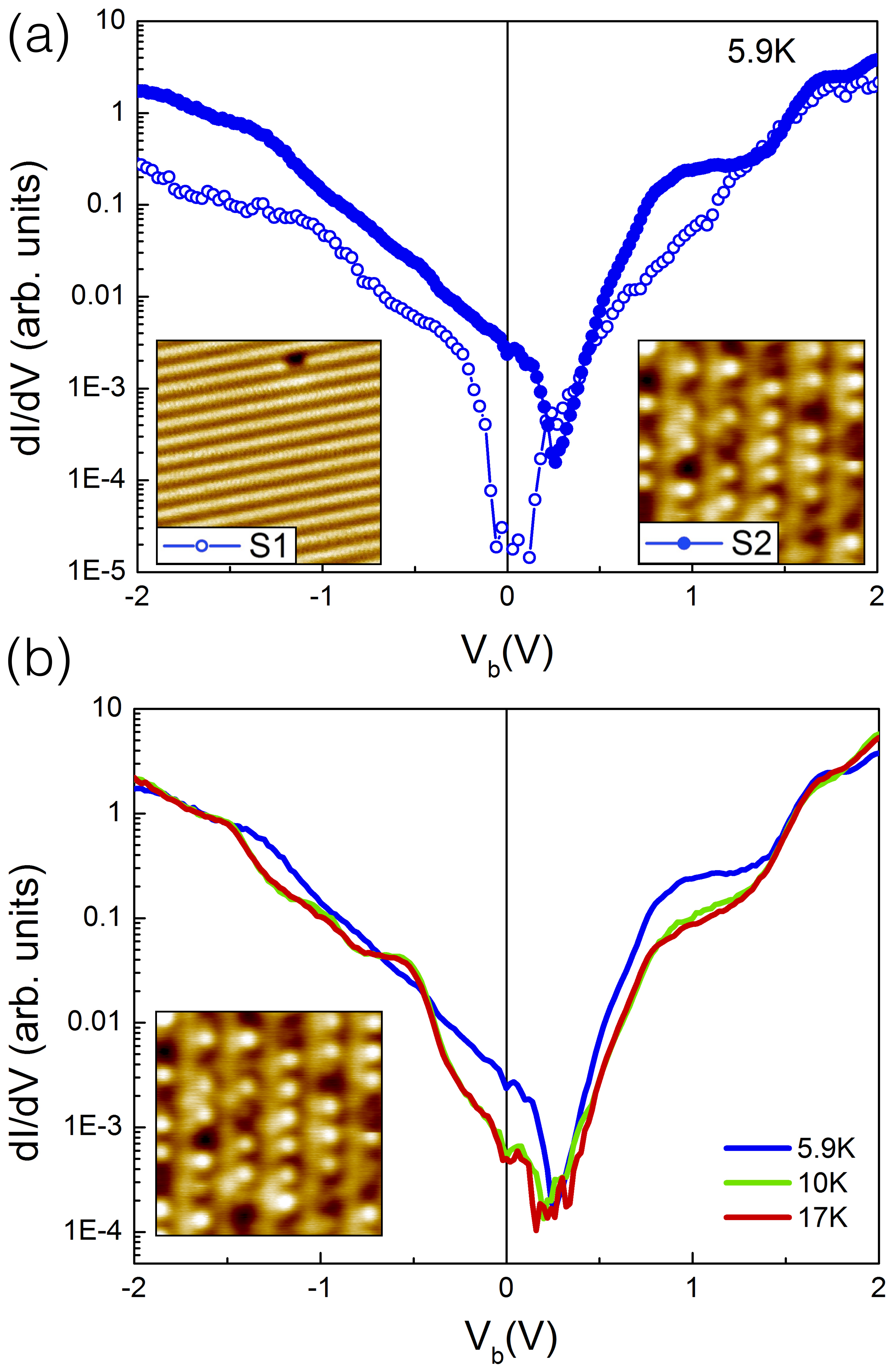}
\caption{(a) Averaged spectra taken at 5.9\,K on two atomically flat surfaces
of \eis: the non-reconstructed $ac$ plane, S1, and the (081) plane, S2. The
insets present the corresponding topographies. (b) $T$-dependence of
d$I/$d$V$-spectra on S2 covering temperatures below, in between and above
$T_{\rm N1}$ and $T_{\rm N2}$.}
\label{FigD.pdf}
\end{figure}
Notably, the density of defects, apart from the step edges, is very low. This arrangement can be nicely correlated to the double-chain stacking, where each
stripe corresponds to the end-part of the chains, see Fig.\ \ref{FigC.pdf}(e),
constituting a Sb-terminated surface. An alternative scenario arises by
considering the surface composed by the Eu(1) and Eu(2) ions and formed by
breaking the Sb dumbbell bonds during the cleaving process, cf.\ Fig.\
\ref{FigC.pdf}(f), again forming a height-modulated surface with periodicity
$a$ and possible step edges of $\frac{1}{2}b$. On both types of terminations,
there is an in-plane displacement between the corrugations on both sides of a
\begin{table*}[t]
\caption{Results of total energies calculated by DFT for different spin
configurations in \eis\ and the spin-moments on the Eu sites of the three
different sublattices. The spin configurations FM, FiM, and AFM represent
ferromagnetic, ferrimagnetic, and antiferromagnetic ones, respectively, as
illustrated in Fig.\ \ref{FigCfgs}(except FM).}
\label{TabEnergiesAFMDFT}
\begin{ruledtabular}
\begin{tabular}{l|c|ccc|r}
Name & Configuration\hspace*{1cm} & \multicolumn{3}{c|}{Magnetic spin moments}
& Energy \\  \hline
& lattice sites\hspace*{1cm} & $|m_s^{(1)}| $ & $|m_s^{(2)}|$
& $|m_s^{(3)}|$ & relative to AFM-a\\
& (Eu(1))(Eu(2))(Eu(3))\hspace*{1cm} & & $[\mu_B/$Eu$]$ & &
$[$meV$/$unit cell$]$ \\
\hline
FM   & $(\uparrow \uparrow \uparrow \uparrow)$$(\uparrow \uparrow)$$(\uparrow \uparrow \uparrow \uparrow)$\hspace*{1cm} & 7.154 & 7.169 & 7.160 & 87.600 \\
FiM & $(\uparrow \downarrow \uparrow \downarrow)$$(\uparrow \uparrow)$$(\uparrow \downarrow \uparrow \downarrow)$\hspace*{1cm} & 7.200 & 7.189 & 7.203 & 2.164\\
AFM-c& $(\uparrow \downarrow \uparrow \downarrow )$$(\uparrow \downarrow)$$
(\uparrow \downarrow \uparrow \downarrow)$\hspace*{1cm} & 7.202 & 7.201 &
7.211 & 0.055 \\
AFM-b& $(\uparrow \downarrow \uparrow \downarrow )$$(\downarrow \uparrow)$$
(\uparrow \downarrow \uparrow \downarrow)$\hspace*{1cm} & 7.209 & 7.201 &
7.206 & 0.055 \\
AFM-a& $(\uparrow \uparrow \downarrow \downarrow)$$(\downarrow \uparrow)$$ (\downarrow \downarrow \uparrow \uparrow)$\hspace*{1cm} & 7.179 & 7.210 &
7.212 &0\\ \end{tabular}
\end{ruledtabular}
\end{table*}
step edge, cf.\ Figs.\ \ref{FigC.pdf}(e) and (f); the expected value in the Sb-terminated scenario is $0.21$\,nm and for the Eu-terminated one $0.19$\,nm.
Unfortunately, the small difference between the values and the lack of atomic
resolution along the $c$ axis renders an identification of the surface
termination challenging. Nonetheless, the small difference of the Eu(1)
positions along $b$ direction and the expected asymmetry of the stripes along
$a$ make an Eu-terminated surface less likely. Also from an energetic point of
view, an Eu-terminated surface appears unfavorable considering the involved
Sb$_2$-Sb$_2$ bond-breaking. Yet, further surface energy calculations are
needed to assess the surface termination.

Finally, motivated by the predictions of topological surface states on the
(001) surface of Ba$_5$In$_2$Sb$_6$ \cite{wie18}, we attempted to cleave
samples \#4 and \#5 along the $c$ axis. Unfortunately and despite exhaustive
efforts, atomically flat surface areas could not be found. Intuitively, the
experimental absence of ordered flat surfaces may be understood by considering
that the breaking of the double chains is very likely unfavorable.

\subsection{STS}
\label{sec:STS}
An overview of the local conductance, $g(V,T) = {\rm d}I(V)/{\rm d}V$, as
measured by STS, is presented in Fig.\ \ref{FigD.pdf}. A comparison of the
results obtained on two atomically flat surfaces at $5.9\,$K is presented in
Fig.\ \ref{FigD.pdf}(a): the non-reconstructed $ac$ surface of sample \#3,
here called S1, and the (081) plane of sample \#1, denoted S2. Albeit $g(V)$
of both spectra span several orders of magnitude (note the logarithmic scale)
within the range $-2$\,V $\leq V_b \leq +2$\,V with deep, yet finite minima,
there are clear differences: i) $g(V)$ is significantly higher on S2,
specifically for $V_b \leq 0$. ii) While the minimum of $g(V)$ is very close
to $V_b = 0$ for S1, it appears to be shifted to $V_b \approx$ 260\,mV. iii)
the gap width appears to  be slightly smaller on S2. Likely, these observations
result from the more disturbed surface with larger defect density of S2, given
the insulating behavior of \eis\ at this low temperature. However, it should
be emphasized that the measured $g(V)$-values remain finite within the pseudo
gap even for the well-ordered surface S1.

The temperature dependence of $g(V,T)$ was studied on surface S2, see Fig.\
\ref{FigD.pdf}(b). In particular, we repeated the measurements below and above
$T_{N1}$ and $T_{N2}$ several times in order to evaluate any possible
influence of the antiferromagnetic order on $g(V,T)$. At all temperatures,
the minimum of $g(V,T)$ is found at positive $V_{\rm b}$. In general, there
is no significant difference between the spectra taken at $10\,$K and $17\,$K,
i.e.\ around $T_{\rm N1} = 14.1$\,K. In contrast, below $T_{\rm N2}= 7.3$\,K
$[$see spectrum at $T=5.9$\,K in Fig.\ \ref{FigD.pdf}(b)$]$ the conductance at
$V_{\rm b} = 0$, and therefore the LDOS at the Fermi level, is increased by
almost one order of magnitude.

\subsection{Density Functional Theory calculation}
\label{sec:DFT}
In order to calculate the DOS of the ground state using DFT, it is necessary
to first determine the ground state magnetic configuration. In consequence,
we start by describing the calculation of energies for the different magnetic
configurations, and subsequently, calculate the DOS of the energetically most
favourable configurations.

\subsubsection{Magnetic configurations}
\begin{figure}[t]
\centering
\includegraphics[clip,width=0.95\columnwidth]{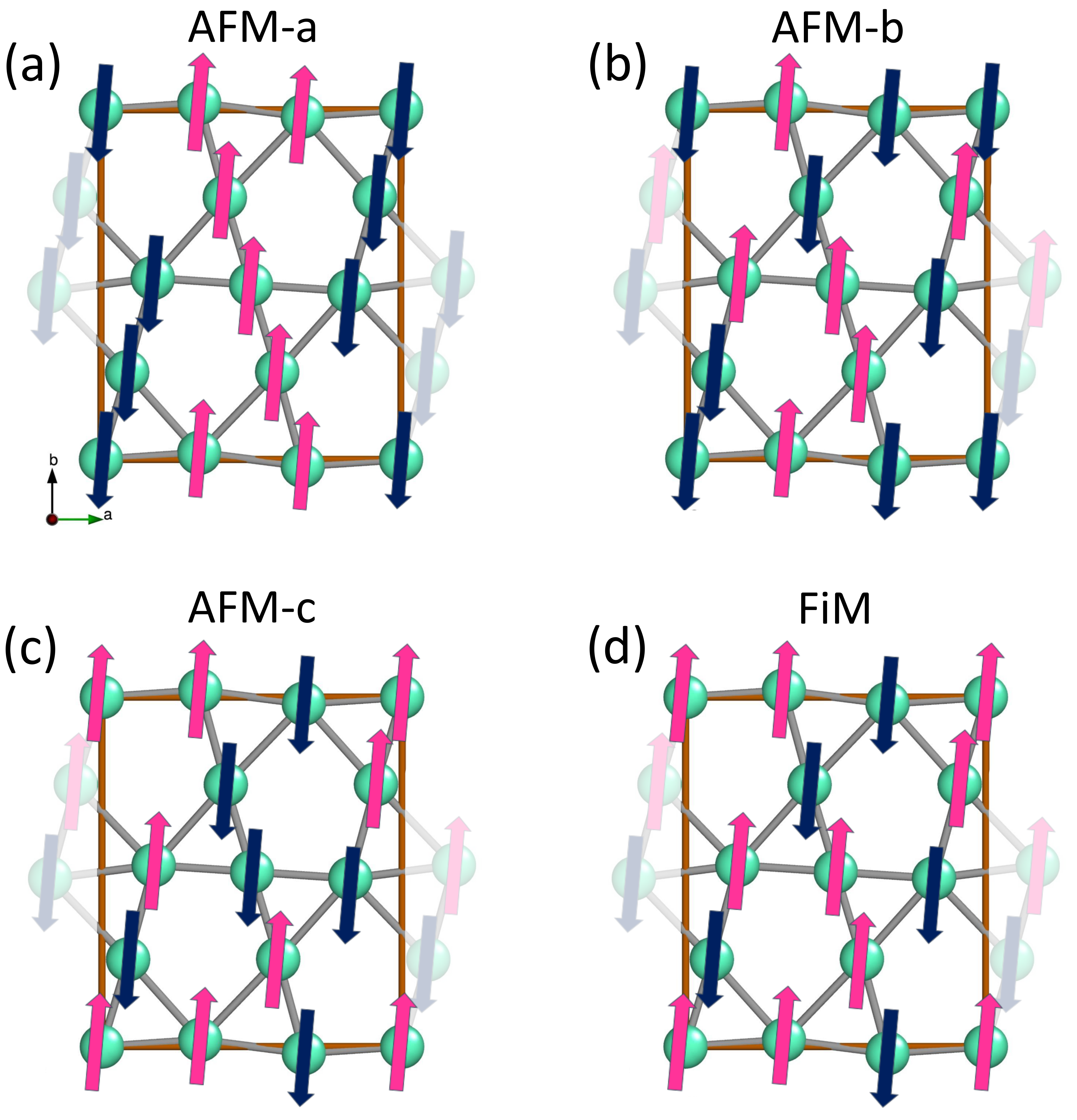}
\caption{Magnetic configurations of lowest energy according to DFT
calculations: (a)--(c) fully compensated antiferromagnetic (AFM) states from
collinear DFT (spin resolved). (d) first ferrimagnetic (FiM) configuration.
The crystallographic orientation conforms to the one in
Fig.\ \ref{FigG.pdf}(b).}  \label{FigCfgs}
\end{figure}

\eis\ contains ten Eu ions in the $ab$ base plane of the unit cell. These sites
belong to three different Wyckoff positions of the space group \textit{Pbam};
Eu(1) and Eu(3) both occupy two different sets of 4g sites and Eu(2) the 2a
sites. In order to probe possible magnetic configurations, we have conducted
DFT calculations for all combined magnetic states that can be created from
either fully saturated ferromagnetic (FM) or fully compensated
antiferromagnetic (AFM) collinear spin-configurations of these three magnetic
sub-systems, amounting to 128 magnetic configurations. The calculations were
executed with the scalar relativistic DFT approach and using the open-core
method. The four configurations with lowest calculated energies are sketched
in Fig.\ \ref{FigCfgs}. The corresponding energies with respect to the AFM-a
state and the magnetic spin-moments are listed in
Table~\ref{TabEnergiesAFMDFT}. The A-type configuration has ferromagnetically
coupled Eu spin moments in the $ab$-plane, that are alternatingly stacked in
an antiparallel stagggered order. Such a spin structure has been proposed as
possible magnetic order in Ref.~\cite{ros20}.

In addition, for the antiferromagnetic configurations with lowest energies,
fully relativistic calculations were conducted. The corresponding results
suggest that: i) The effective coupling between the Eu-layers likely is
ferromagnetic. ii) There are strong antiferromagnetic couplings within the
$ab$-plane between Eu-ions within the three sublattices of Eu(1)-, Eu(2)-
and Eu(3)-sites. iii) The close match of energies for the different AFM and
the ferrimagnetic configuration indicates that this multi-sublattice spin
system is likely frustrated. The reason is presumably geometric frustration
related to triangular and pentagonal elementary plaquettes between
nearest-neighbor Eu-sites. In consequence, the correct magnetic ground state
of \eis\ is likely a complex non-collinear AFM or ferrimagnetic state to lift
this frustration. However, magnetic couplings beyond next neighbors may also
play a role. The magnetic exchange mechanism between the Eu$^{2+}$-ions is
likely of the Ruderman-Kittel-Kasuya-Yosida (RKKY)-type which can support
long-range contributions, both in-plane and out-of-plane. Our restricted
computational results do not allow a complete computational determination of
the magnetic ground state which includes these intricacies.

It is noteworthy that the lowest energy magnetic configurations do not break
inversion symmetry and, if no cell doubling is enforced by these long-range
\begin{figure}[t]
\centering
\includegraphics[clip,width=0.95\columnwidth]{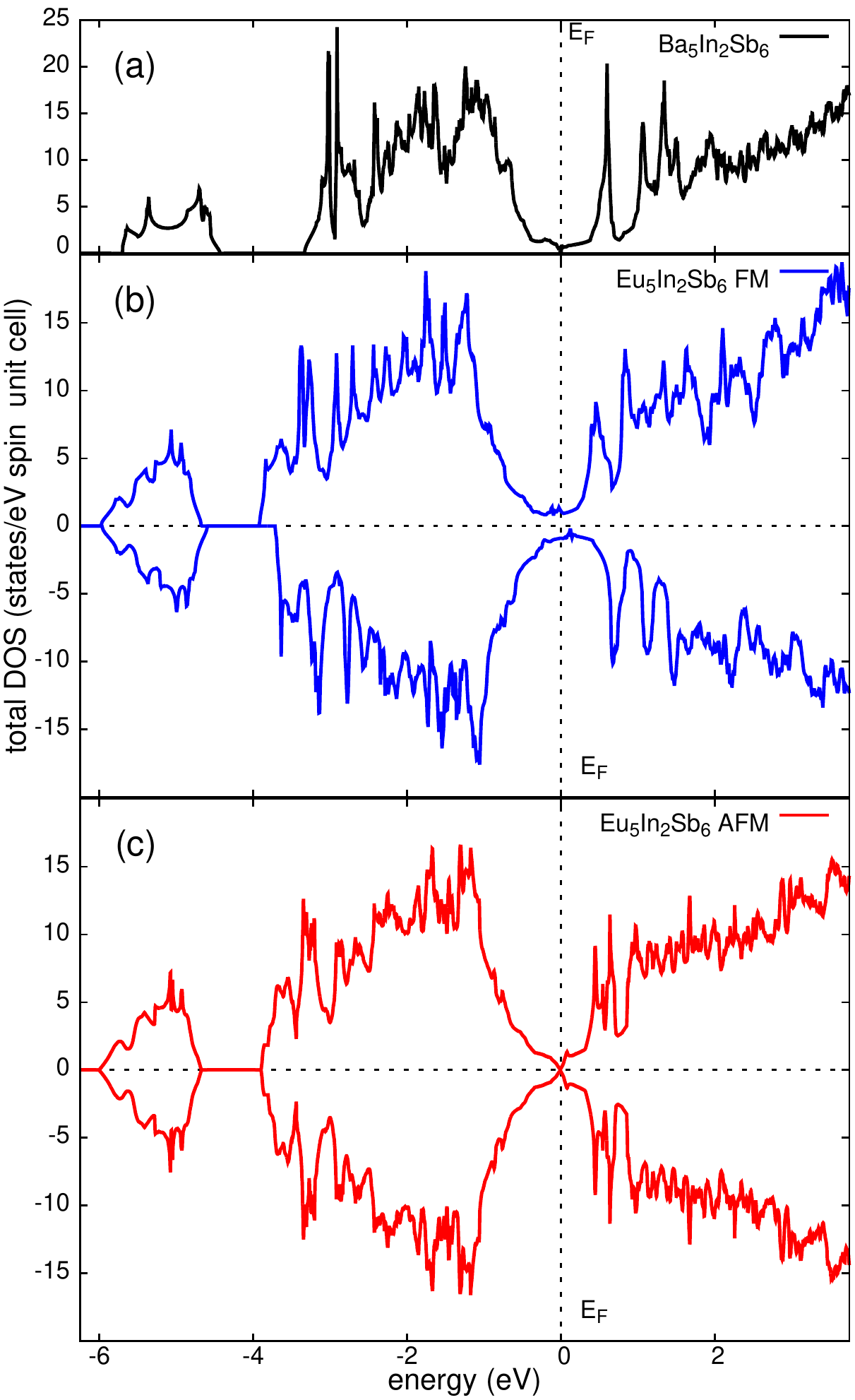}
\caption{Density of states (DOS) from scalar relativistic (sr) DFT
calculations: (a) Ba$_{5}$In$_{2}$Sb$_{6}$, (b) \eis\ in the ferromagnetic
(FM) state. (c) \eis\ in antiferromagnetic state, AFM-a configuration.}
\label{FigDOS}
\end{figure}
couplings, the system of magnetic $\Gamma$-point modes calculated is complete
in the so-called exchange approximation. If the magnetic ground-state of \eis\
is described by a suitable combination of such $\Gamma$-point modes, a
relatively simple spin structure with compensated moments may approximately
describe this magnetic system. However, in that case spin-orbit coupling (SOC)
does allow for the occurrence of weak ferromagnetism by spin-canting in this
orthorhombic magnetic crystal \cite{bog02}, as the Dzyaloshinskii-Moriya
exchange is allowed. In fact, experimental data on \eis\ revealed a very small
in-plane ferromagnetic component and hysteresis which was ascribed to the
complex magnetic structure with canted moments \cite{ros20}. As such, the
intrinsic weak ferromagnetism may be linked to the primary AFM $\Gamma$-point
order.

Sizeable effects by SOC are seen in the results of fully relativistic (fr)
calculations with different quantization axes in Table~\ref{TabEnergiesAFMDFT}.
From an experimental point of view, a determination of the magnetic ground
state and a detailed understanding of the magnetic field--temperature phase
diagram for \eis\ will require major efforts. Our theoretical results, however,
suggest that the basic spin structure is antiferomagnetic within the $ab$
plane. A simple A-type antiferromagnetic structure appears unlikely.
Nonetheless, the evaluation of the electronic band structure below is based on
the collinear low-energy spin structures (Fig.\ \ref{FigCfgs}) which are
expected to approximate the essential features of ordered magnetism in \eis.
\subsubsection{Electronic band structure}
In an effort to gain further insight into the band structure of \eis\ and to
allow for comparison to our STS results (Fig.\ \ref{FigD.pdf}), the electronic
density of states (DOS) for both the ferromagnetic (FM) and the AFM spin
configuration AFM-a with lowest energy were calculated, see Fig.~\ref{FigDOS}.
In addition, the DOS of the non-magnetic isostructural compound
Ba$_{5}$In$_{2}$Sb$_{6}$ is shown. All three systems behave as semi-metals
\begin{figure}[t]
\centering
\includegraphics[width=0.96\columnwidth]{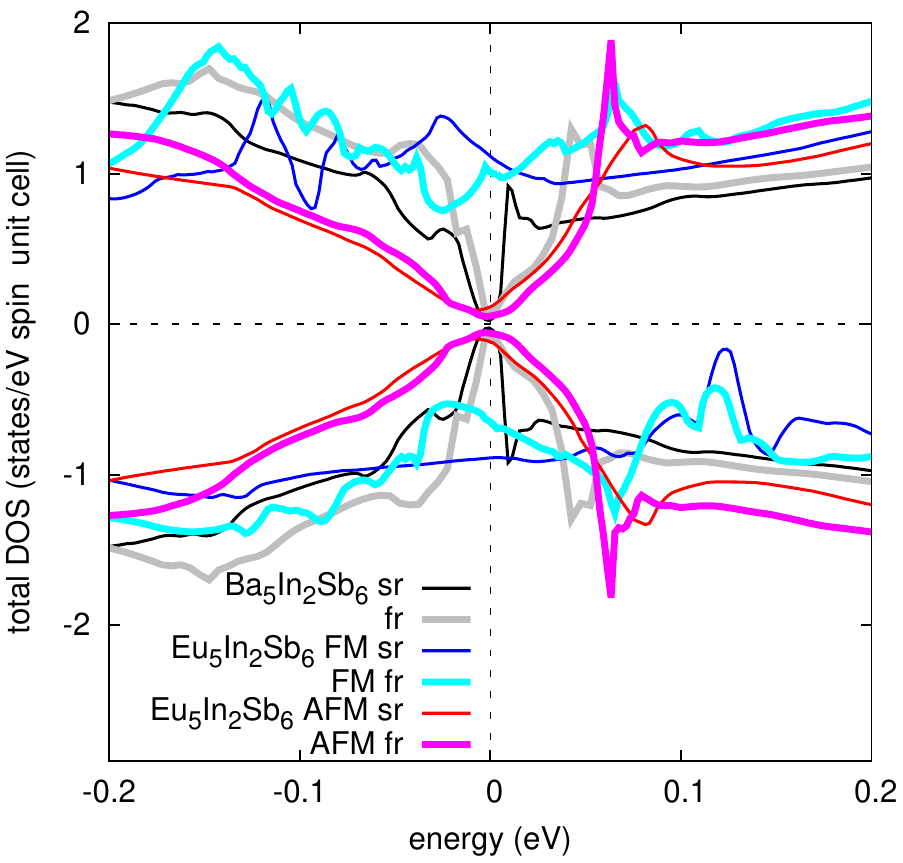}
\caption{DOS near the Fermi energy $E_{\rm F}$ (at zero energy). A comparison
between fully relativistic (fr) and scalar relativistic (sr) DFT calculations
for \eis\ in FM and AFM spin configuration as well as Ba$_{5}$In$_{2}$Sb$_{6}$
is shown.} \label{FigDosNearEF}
\end{figure}
with a small DOS near the Fermi energy $E_{\rm F}$. Qualitatively, the
gap-like features observed for \eis\ in both the calculations and the STS data
even share the asymmetry of a somewhat steeper rise for positive energies
compared to the negative side.

A closer inspection of the energy region near $E_{\rm F}$ (Fig.\
\ref{FigDosNearEF}) reveals a significant difference between the FM and the AFM
configuration. The AFM state with compensated total moments has a significant
depression of the DOS right at $E_{\rm F}$ compared to the FM state. These
results for the AFM state are also in line with the STS data discussed above.
The comparison of fully-relativistic (fr) and scalar relativistic band
structure calculations near $E_{\rm F}$ also shows a remarkable re-distribution
and more structured DOS for the former, but it does not change the overall
picture.

In an effort to gain insight into a possible nontrivial electronic band
topology of \eis, the momentum-resolved band structure as obtained by a
\begin{figure}[t]
\centering
\includegraphics[width=0.96\columnwidth]{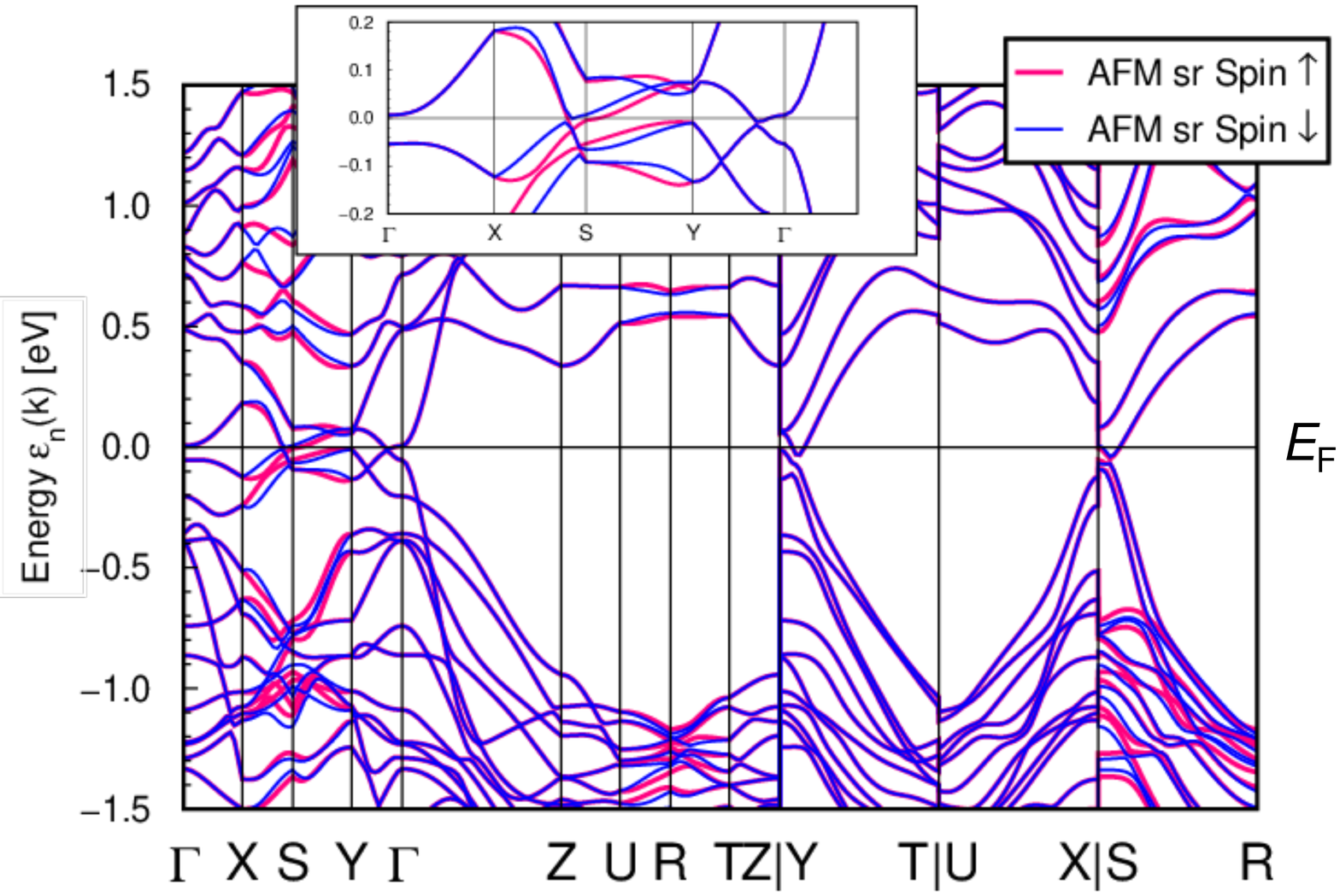}
\caption{Band structure of \eis\ in the AFM-a spin conﬁguration. The insets
show details near the Fermi energy $E_{\rm F}$, which is set to zero energy.
The scalar-relativistic DFT calculations are represented by spin-split band
data for spin majority and minority components.} \label{bandsAFM}
\end{figure}
scalar relativistic DFT calculation for AFM-a configuration is presented in
Fig.\ \ref{bandsAFM}. Clearly, there are certain spin-split bands crossing
$E_{\rm F}$ in the AFM state, see section along X--S--Y in the energy-enlarged
inset. These results are in good agreement with those obtained by other band
structure calculations \cite{yfxu}. It should be noted, however, that care has
to be taken when comparing results of different calculation schemes: On the
one hand, the magnetism of \eis\ is included in our calculations. On the other
hand, SOC is only considered in the fully relativistic calculations presented
in Fig.\ \ref{FigDosNearEF} while it is neglected in the scalar relativistic
ones. Our calculations do not provide any indication for a non-trivial band
topology, in line with a trivial $\mathbb{Z}_2$ index proposed in Ref.\
\cite{var22}. This is consistent with our STS data, at least for the surfaces
investigated so far. Nonetheless, to scrutinize the surface topology in \eis,
the topological invariants for the double-glide (001) surface should be
explored by Wilson loop calculations, similar to those for Ba$_5$In$_2$Sb$_6$
in Ref.\ \onlinecite{wie18}.

\section{Discussion and conclusions}
\label{sec:Discussion}
\eis\ was successfully cleaved \textit{in situ} and patches of atomically flat
areas, specifically for the $ac$ surface plane, were located. Within such
areas, our STS data indicated gap-like spectra with a very low, but finite
conductance at the Fermi level. Qualitatively, the STS data agree with the
calculated DOS in the energy range from $-1$ to $+1$ eV. The calculated band
structures and resultant DOS for the AFM and FM spin structures in \eis\
ultimately illustrate how the difference in spin configuration can lead to a
re-organization of the small band contributions near $E_{\rm F}$. Apart from
such subtlety, the major influence of the underlying Eu 4$f$ spin structure
suggests a rather traditional picture for the electronic properties of this
compound. In the AFM state, the material has only rather few charge carriers
available, and the AFM background allows only for incoherent hopping transport
via thermal activation, which may lead to an effective insulator-like behavior
as indeed observed experimentally. In particular, a negative temperature
coefficient of the resistivity, $d\rho /dT < 0$ may be caused by the stiffening
of the AFM correlations upon lowering the temperature and even through the
magnetic phase transitions. For decreasing temperature and/or under
applied magnetic field, spin-polarons can be formed and may lead to strongly
improved conduction, resulting in a CMR behavior. The enhanced DOS in the FM
state shows $[$see Fig.\ \ref{FigDOS}(b)$]$, how such a spin-polarized
electronic structure generates charge carriers.

At present, neither our band structure calculations for the AFM-a state nor
the STS results at low temperature provide any indication for a non-trivial
band topology of \eis. Nonetheless, our future experimental efforts will focus
on an optimization of cleaved (001) surfaces as atomically flat surfaces of
this orientation appear to be essential for conclusive results concerning the
topological nature of \eis.

\begin{acknowledgments}
We thank U. Nitzsche for technical assistance with respect to the DFT
calculations. Work at Los Alamos was supported by the Los Alamos Laboratory
Directed Research and Development program through project 20220135DR. SR
acknowledges support by the Deutsche Forschungsgemeinschaft (DFG, German
Research Foundation) through SFB 1143. Work at the Max-Planck-Institute for
Chemical Physics of Solids in Dresden and at Goethe University Frankfurt was
funded by the Deutsche Forschungsgemeinschaft (DFG, German Research
Foundation), Project No.\ 449866704.
\end{acknowledgments}

\end{document}